\documentclass[aps,prd,superscriptaddress]{revtex4}

\usepackage{amsfonts}
\usepackage{amsmath}
\usepackage{graphicx}
\usepackage{color}
\usepackage{amssymb}
\usepackage{esint}
\usepackage{empheq}
\usepackage{mathtools,leftidx}
\usepackage{float}

\usepackage{soul}

\DeclareMathAlphabet\mathbfcal{OMS}{cmsy}{b}{n}

\begin{document}

\title{ Lorentz violating scalar Casimir effect for a $D$-dimensional sphere}

\author{A. Mart\'{i}n-Ruiz}
\email{alberto.martin@nucleares.unam.mx}
\affiliation{Instituto de Ciencias Nucleares, Universidad Nacional Aut\'{o}noma de M\'{e}xico, 04510 Ciudad de M\'{e}xico, M\'{e}xico}

\author{C. A. Escobar}
\email{carlos_escobar@fisica.unam.mx}
\affiliation{Instituto de F\'{i}sica, Universidad Nacional Aut\'{o}noma de M\'{e}xico, Apartado Postal 20-364, Ciudad de M\'{e}xico 01000, M\'{e}xico}

\author{A. M. Escobar-Ruiz}
\email{admau@xanum.uam.mx}
\affiliation{Departamento  de  F\'isica,  Universidad  Aut\'onoma  Metropolitana-Iztapalapa, San Rafael Atlixco 186, 09340 Ciudad de M\'{e}xico, M\'{e}xico}

\author{O. J. Franca}
\email{francamentesantiago@ciencias.unam.mx}
\affiliation{Instituto de Ciencias Nucleares, Universidad Nacional Aut\'{o}noma de M\'{e}xico, 04510 Ciudad de M\'{e}xico, M\'{e}xico}

\begin{abstract}
We investigate the Casimir effect, due to the confinement of a scalar field in a $D$-dimensional sphere, with Lorentz symmetry breaking. The Lorentz-violating part of the theory is described by an additional term $\lambda (u \cdot \partial \phi) ^{2}$ in the scalar field Lagrangian, where the parameter $\lambda$ and the background vector $u^{\mu}$ codify the breakdown of Lorentz symmetry. We compute, as a function of $D>2$, the Casimir stress by using Green's function techniques for two specific choices of the vector $u ^{\mu}$. In the timelike case, $u ^{\mu} = (1,0,...,0)$, the Casimir stress can be factorized as the product of the Lorentz invariant result times the factor $(1 + \lambda) ^{-1/2}$. For the radial spacelike case, $u ^{\mu} = (0,1,0,...,0)$, we obtain an analytical expression for the Casimir stress which nevertheless does not admit a factorization in terms of the Lorentz invariant result. For the radial spacelike case we find that there exists a critical value $\lambda _{c} = \lambda _{c} (D)$ at which the Casimir stress transits from a repulsive behavior to an attractive one for any $D> 2$. The physically relevant case $D = 3$ is analyzed in detail where the critical value $\lambda _{c}|_{\small D=3} = 0.0025$ was found. As in the Lorentz symmetric case, the force maintains the divergent behavior at positive even integer values of $D$.
\end{abstract}

\maketitle

\section{Introduction}

Observable macroscopic forces produced by quantum vacuum fluctuations of the electromagnetic field have attracted great attention in theoretical and experimental studies. The most renowned are perhaps the attractive Casimir force between two neutral bodies  \cite{Casimir,Milonni,Dalvit-Milonni-Roberts-Rosa,Woods-et-al} and the Casimir-Polder force between an atom and a neutral body \cite{Casimir-Polder}. The Casimir effect, for example, has found various applications in macroscopic physics, cosmology, hadron physics, supersymmetry and supergravitation. Since the Casimir effect (CE) has been verified with an astonishing high precision \cite{Lamoreaux,Mohideen,Roy,Klimchitskaya}, it has become a testing ground for the predictions of new fundamental physical theories.

Outside the paradigmatic CE between two parallel conductive plates, many theoretical and experimental researches have been conducted regarding the calculation of the Casimir force for different geometries and spacetime topologies \cite{Deutsch-Candelas,DeRaad-Milton,Dowker, Aliev,Huang}. In particular, the CE in the cylindrical and spherical geometries have attracted the most attention due to its simplicity and important applications. Interestingly, the former leads to an attractive (although small) force \cite{DeRaad-Milton} while the latter produces a repulsive force \cite{Boyer}. Regarding the applications, the CE for a sphere has proved to be useful in the understanding of other phenomena. For example, the van der Waals force is equivalent to the CE for a dielectric sphere \cite{Milton 1, Milton's book}, and it has also been considered as the responsible for the production of visible-light photons in the bubble collapse which occurs in sonoluminescence \cite{Milton 2}. It has also turned out to be strikingly important in hadron physics, particularly in the context of the bag model and chiral bag model \cite{Johnson}. In these systems, quarks and gluons are absolutely confined inside the bag which is bounding a hadron, and hence the Casimir energy of these fields must be incorporated in the total energy of a bag in hadron properties calculations.

The analysis of the CE and other physical systems in $D$ spatial dimensions have been found useful. In quantum field theory, for example, spacetime dimensions have been employed as a perturbative parameter, thus allowing to provide analytical solutions that are nonperturbative in the coupling constant. On the other hand, the CE for a $D$-dimensional sphere has also been used in hadron physics. As shown in Ref. \cite{F-G-K}, the zero-point energy in the flux-tube connecting a heavy quark-antiquark system can be regarded as the zero-point energy of an hypersphere with dimension $D=1$. Also, in the same context, it has been applied to study the quantum stabilization of $1+1$-dimensional static solids \cite{Graham 1} and  the quantum energies of interfaces \cite{Graham 2, Graham 3}. These studies exemplify the usefulness of the CE in $D$ spatial dimensions in the context of hadron physics, and suggest the need for additional research in this direction. This is precisely the main goal of this work, but within the context of a Lorentz-violating (LV) scalar field theory.

Lorentz violation is currently a topic of great interest in particle physics. This is motivated from the fact that the standard model, although phenomenologically successful, suffers from some theoretical inconsistencies and long-standing unresolved problems. Nowadays, investigations concerning Lorentz violation are mostly carried out under the framework of the Standard-Model Extension \cite{SME2,Kostelecky 3}, which contains  translation-invariant but Lorentz-violating corrections to the standard model parameterized by small tensor-valued background fields. No evidence for a deviation from Lorentz invariance has been found, but experimental Lorentz tests are constantly being refined. Due to its potential scope, the CE stands as a good arena to test Lorentz-violating field theories. This has motivated the study of the CE in Lorentz-violating scenarios in the classical geometry of two parallel plates  \cite{Cruz 1, Cruz 2, Escobar-Medel-Martin, PLB}. This, together with the above discussed applications of the CE for a sphere, motivate the present work, where we study how Lorentz symmetry violation manifests on the CE for a real scalar field when it is confined to a $D$-dimensional spherical shell.

The outline of this paper is as follows. Section \ref{LV Sec} introduces the theoretical model for a real massive scalar quantum field $\phi$ in the presence of Lorentz violation. This consists of the Klein-Gordon Lagrangian density supplemented with the Lorentz-violating term $\lambda\left(u\cdot\partial\phi\right)^2$, where $\lambda$ is a parameter and $u^{\mu}=(u^0,\vec{u})$ is a constant vector which control Lorentz symmetry breaking \cite{Gomes-Petrov}. As required by the local approach to the CE, in Sec. \ref{GFsection} we derive the Green's function for the two cases we consider in the present study: the timelike case for which $u ^{\mu} = (1,0, ... , 0)$ and the radial spacelike one where $u ^{\mu} = (0,1,0 ... , 0)$. In Sec. \ref{CEsection} we derive analytical expressions for the Casimir stress in each case. Afterwards, in Sec. \ref{Numerical Section} we follow the procedure reported in Ref. \cite{Bender-Milton} to evaluate numerically the Casimir force in the massless case and we present the results in Sec. \ref{NumResultsSec}. Finally, Sec. \ref{ConcluSection} summarizes our results and gives further concluding remarks. Throughout the paper, natural units are assumed $\hbar = c = 1$.

\section{General settings} \label{LV Sec}

Let us consider the LV Lagrangian density for a massive real scalar field \cite{Cruz 1}
\begin{align}
\mathcal{L}  = \frac{1}{2} \left[ \partial _{\mu} \phi \, \partial ^{\mu} \phi +  \lambda \left( u ^{\mu} \, \partial _{\mu} \phi \right) ^{2} - m ^{2} \, \phi ^{2}  \right], \quad \quad \quad \mu=0,1,2,...,D \; ,
\label{Lagrangian1}
\end{align}
where $u ^{\mu} \equiv (u _{0} , \vec{u} \, )$ denotes a non-zero $(D+1)$-dimensional constant vector and $ \vert \lambda \vert < 1$. The second term in Eq. (\ref{Lagrangian1}) mimics a background field that specifies privileged directions on the spacetime and encodes the Lorentz symmetry violation.
In this work we assume that the scalar field $\phi(x)$ is confined in a $D$-dimensional sphere of radius $R$. Formally, one
can think $D$ as a continuous variable that ranges from $0$ to $\infty$. In the present study we restrict ourselves to the case $D>2$ only. The Lorentz-invariant case corresponds to $\lambda=0$. It is worth mentioning that the LV term $\lambda \left( u ^{\mu} \partial _{\mu} \phi \right) ^{2}$ was originally conceived within the scalar sector of the Standard-Model Extension (SME) \cite{Kostelecky 3,SME2}. In the SME context, it is expected that physical relevant systems correspond to $\vert  \lambda u ^{\mu} u ^{\nu} \vert \ll 1$ for any $\mu$ and $\nu$. However, there are scenarios in which Lorentz symmetry is naturally broken, such as condensed matter systems, where Lorentz-violating coefficients are not much smaller than one. Examples of this are properly discussed in Sec. \ref{NumResultsSec}. These Lorentz-violating models also appear in connection to Riemann-Finsler spacetimes \cite{Finsler} where the notion of distance is controlled by additional quantities beyond the Riemann metric, which can intuitively play the role of Lorentz-violating coefficients \cite{KFinsler}. The model in the present work is a particular case of those presented in Ref. \cite{Finsler} through the identification $(\hat{k}_c)^{\mu\nu}=\lambda u^\mu\, u^\nu$.

From the Lagrangian (\ref{Lagrangian1}) it follows the modified Klein-Gordon equation
\begin{align}
\left[ \, \Box \ + \ \lambda  \left( u _{\mu} \, \partial ^{\mu} \right) ^{2} \ + \ m ^{2} \right] \,\phi (x) \  = \ 0 , \label{EQLV}
\end{align}
where the symbol $\Box = \partial _{\mu} \partial ^{\mu}$ stands for the standard D'Alembert operator in $(D+1)$-dimensions. We will consider solutions of the equation of motion (\ref{EQLV}) that satisfy Dirichlet boundary conditions (BCs) on the sphere. The corresponding stress-energy tensor takes the form
\begin{align}
T ^{\mu \nu} \ = \ (\partial ^{\mu} \phi ) (\partial ^{\nu} \phi ) \ + \ \lambda  u ^{\mu} \, (\partial ^{\nu} \phi ) (u _{\sigma} \partial ^{\sigma} \phi ) \ - \ \eta ^{\mu \nu} \mathcal{L}\ , \label{Tmunu}
\end{align}
where $\eta ^{\mu \nu} = \textrm{diag} (1,-1,-1,...,-1)$ is the usual Minkowski flat spacetime metric in $(D+1)$-dimensions. It can be checked that the stress-energy tensor (\ref{Tmunu}) is conserved, i.e. $\partial _{\mu} T ^{\mu \nu } = 0$. However, it is not traceless $T ^{\mu} _{\phantom{\mu} \mu} \neq 0$ and, unlike most of the cases where Lorentz symmetry is preserved, it cannot be symmetrized \cite{Cruz 1}.

Since the spherical shell divides the space into two regions, the interior and exterior of the sphere, it is natural to use hyperspherical coordinates for the spacelike components of the stress-energy tensor (\ref{Tmunu}). In these coordinates, the Casimir force per unit area $F/A$ on the sphere is obtained from the discontinuity of the radial-radial component of the vacuum expectation value of the stress-energy tensor \cite{Milton's book}
\begin{align}
\frac{F}{A} = \langle \, 0 \, \vert \, T ^{rr}  _{\mbox{\scriptsize in}} - T ^{rr}  _{\mbox{\scriptsize out}} \, \vert \, 0 \, \rangle \big\vert_{\| \vec{x}\|\,=\,R}  \,  , \label{F1a}
\end{align}
where the subindices in/out indicate the region where the vacuum expectation value of $T ^{rr}$ has to be evaluated.

In the present study two particular cases will be analyzed in detail, namely
\begin{itemize}
  \item[(I)] the radial spacelike case $u ^{\mu} = (0,1,0, \cdots ,0)$, where all the coordinates of $u^\mu$ are zero except the radial spacelike coordinate, and

  \item[(II)] the timelike case for which $u ^{\mu} = (1,0, \cdots ,0 )$, where the timelike component $u _{0}$ is different from zero only.
\end{itemize}
Equivalently, in terms of the two-point Green's function (GF) defined as the vacuum expectation value of the time-ordered product of two fields \cite{Peskin}
\begin{align}
G(x , x ^{\prime} ) = - i \,\langle\, 0 \, \vert \, \hat{\mathcal{T}} \phi(x) \phi( x ^\prime) \, \vert 0 \, \rangle , \label{Grel}
\end{align}
the radial Casimir force (\ref{F1a}) can be rewritten as 
\begin{align}
\frac{F}{A} = - \frac{i}{2} \, \Lambda \lim _{x ^{\prime} \rightarrow {x}} \bigg[ \frac{\partial }{\partial r}\frac{\partial}{\partial r ^{\prime}} G _{\mbox{\scriptsize in}} (x,x ^{\prime}) - \frac{\partial }{\partial r} \frac{\partial}{\partial r ^{\prime}} G _{\mbox{\scriptsize out}} (x,x ^{\prime}) \bigg] \bigg\vert _{\| \vec{x}\|\,=\,R}  ,
\label{FAbyG}
\end{align}
where $G _{\mbox{\scriptsize in}}$ and $G _{\mbox{\scriptsize out}}$ are the GF for the interior ($r<R$) and exterior ($r>R$) of the sphere, satisfying Dirichlet BC on the surface ($r = R$). Further, $\Lambda = 1 - \lambda$ for case I while $\Lambda = 1$ for case II. In general, there are other terms not included in Eq. (\ref{FAbyG}) which depend on angular derivatives of the GFs that vanishes by virtue of the Dirichlet BC on the sphere.

\section{Green's function}\label{GFsection}

In this section we derive, for the cases I and II, the GF appearing in Eq. (\ref{FAbyG}). As mentioned above, we assume that the GF obeys the Dirichlet BC at the surface of the sphere
\begin{align}
G( x , x ^\prime ) \, \big| _{\| \vec{x}\| = R} = 0 , \label{Dirichlet}
\end{align}
and finiteness at the origin
\begin{align}
G(x,x^\prime) \, \big| _{\| \vec{x}\| = 0} < \infty . \label{Finiteness}
\end{align}
From Eq. (\ref{EQLV}) follows that the corresponding GF satisfies the equation
\begin{align}
\big[ \, \Box +  \lambda \left( u _{\mu} \, \partial ^{\mu} \right) ^{2} +   m ^{2} \, \big] \, G( x , x ^{\prime})\ = \ - \delta ^{(D+1)} (x - x ^{\prime}) . \label{GF eq}
\end{align}
Now, since the GF is translationally invariant in time let us take the time Fourier transform of $G( x , x ^{\prime} )$:
\begin{align}
G _{\omega} ( \vec{x} , \vec{x} ^{\, \prime} ) = \int _{- \infty} ^{\infty} \, dt \, e ^{-i \,\omega (t-t ^{\prime})}\, G( x , x ^{\prime})\ . \label{GFT}
\end{align}
Substituting Eq. (\ref{GFT}) into Eq. (\ref{GF eq}) we find that the reduced GF $G _{\omega} ( \vec{x} , \vec{x} ^{\, \prime} )$ satisfies the differential equation
\begin{align}
\big[ \, \omega ^{2} + \vec{\nabla} ^{2} - \lambda \, (i u _{0} \omega - \vec{u} \cdot \vec{\nabla} ) ^{2} - m ^{2} \, \big] \, G _{\omega} ( \vec{x} , \vec{x} ^{\, \prime} )  =  \delta^{(D)} (\vec{x} - \vec{x} ^{\, \prime} ) .  \label{GF eq omega}
\end{align}
Following the same nomenclature used in Eq. (\ref{FAbyG}), we denote the reduced GFs in the interior and exterior of the sphere as $G _{\omega} ^{(\mbox{\scriptsize in})}$ and $G _{\omega} ^{(\mbox{\scriptsize out})}$, respectively. Also, note that the expression for the Casimir pressure (\ref{FAbyG}) requires the GFs in spacetime coordinates, $G _{\mbox{\scriptsize in/out}} (x,x ^{\prime})$, and this can be obtained by inverse Fourier transforming the reduced GFs $G _{\omega} ^{(\mbox{\scriptsize in/out})} ( \vec{x} , \vec{x} ^{\, \prime} ) $, i.e.
\begin{align}
G _{\mbox{\scriptsize in/out}} (x,x ^{\prime}) = \int _{- \infty} ^{\infty} \, \frac{d\omega}{2 \pi} \, e ^{i \,\omega (t-t ^{\prime})} \, G _{\omega} ^{(\mbox{\scriptsize in/out})} ( \vec{x} , \vec{x} ^{\, \prime} ) . \label{GFT-inv}
\end{align}
Therefore, the problem now consists in determining the reduced GFs.

\subsection{Green's function: radial spacelike case}
\label{GFRa}

First, let us consider the radial spacelike case $u ^{\mu} = (0,1,0,\cdots,0)$. To solve Eq. (\ref{GF eq omega}) it is convenient to introduce polar variables $(r,\theta)$, with $\| \vec{x} \| = r$ and $\theta = \angle(\vec{x},\vec{x} ^{\prime})$ \cite{Bender-Milton}. In terms of these variables, Eq. (\ref{GF eq omega}) becomes
\begin{align}
 \bigg[\omega ^{2} - m ^{2} + \Lambda_R \frac{\partial ^{2}}{\partial r ^{2}} + \frac{D-1}{r} \frac{\partial}{\partial r} + \frac{\sin ^{2-D} \theta}{r ^{2}} \frac{\partial}{\partial \theta} \left( \sin ^{D-2} \theta  \frac{\partial}{\partial \theta}  \right) \bigg] G _{\omega} (r,r ^{\prime} , \theta) = \frac{\delta (r-r ^{\prime}) \delta (\theta) \Gamma \left( \frac{D-1}{2} \right)}{2 \pi ^{\frac{D-1}{2}} r ^{D-1} \sin ^{D-2} \theta } , \label{GFLV ur}
\end{align}
where
\begin{align}
\Lambda _{R} \equiv  1 - \lambda > 0 \ ,
\end{align}
is the effective parameter for Lorentz violation.

As in the Lorentz invariant case ($\Lambda_R=1$), Eq. (\ref{GFLV ur}) admits separation of variables. Its solution can be factorized as
\begin{align}
\label{Godef}
G _{\omega} (r,r ^{\prime} , \theta) = g ( r ,\, r ^{\prime} ) \, \Theta ( \theta ) .
\end{align}
Eventually, we arrive to the following equations
\begin{align}
\left[ -\Lambda_R\, \frac{d ^{2}}{dr ^{2}}\, -\, \frac{D-1}{r} \frac{d}{dr}  \,+\, \frac{n(n+D-2)}{r ^{2}} \,-\, \omega ^{2}\, +\, m ^{2} \right]\, g(r,r ^{\prime}) = 0 , \quad (r \neq r^{\prime}) \label{radial1a}
\end{align}
and
\begin{align}
\left[ \sin ^{2-D} \theta \frac{\partial}{\partial \theta} \left( \sin ^{D-2} \theta  \frac{\partial}{\partial \theta}  \right) + n(n+D-2) \right] \Theta(\theta) = 0 , \label{Angular eq}
\end{align}
where $n$ is a non-negative integer parameter and $n(n+D-2)$ plays the role of the separation constant. We observe that the differential equation (\ref{radial1a}) for the radial part in (\ref{Godef}) depends on the coefficient $\Lambda_R$ for Lorentz violation, while the angular part (\ref{Angular eq}) does not depend on it explicitly. The following remarks are in order:

\begin{itemize}
  \item Only for $D=1$ with $n=0,1$ the second and third terms in the l.h.s of (\ref{radial1a}) vanish simultaneously. In this case, making the substitutions $\omega\rightarrow \omega\,\sqrt{\Lambda_R}$ and $m\rightarrow m\,\sqrt{\Lambda_R}$ in Eq. (\ref{radial1a}), all $\lambda$-dependence is removed and we arrive to the corresponding Lorentz symmetric equation ($\Lambda_R=1$). This special case will be treated separately later on.
  \item In Eq. (\ref{radial1a}), the singular term $\propto r^{-2}$ vanishes at $n=0$ independently of $D$. For any value of $n$ and $D>2$, this term is positive whereas  for $n=1$ and $0<D<1$ it becomes negative. This fact, will be relevant in the construction of the corresponding Green function (see below).
\end{itemize}

Now, the solution to Eq. (\ref{radial1a}) that is well-behaved at the origin is given by
\begin{align}
g _{n} (r,r ^{\prime}) = r ^{\alpha} \left[ \,a _{n} \, J _{s _{n}} (\Omega r) + b _{n} \, Y _{s _{n}} (\Omega r) \,\right]  , \label{gnylv}
\end{align}
where $J_{s _{n}}$ and $Y _{s _{n}}$ are the Bessel functions of the first and second kind \cite{Abramowitz}, respectively, and
\begin{align}
\alpha & = \frac{\Lambda _{R} + 1 - D}{2 \,\Lambda _{R}} , \quad \Omega = \sqrt{\frac{\omega ^{2} - m ^{2}}{\Lambda _{R}}} , \quad s _{n} = \frac{1}{2 \, \Lambda _{R}} \sqrt{ (\Lambda _{R} - D + 1 ) ^{2} + 4 n \Lambda _{R} (n + D -2) } , \label{Coef1}
\end{align}
where we have assumed that $D>2$. This condition will be justified below.

\begin{figure}[H]
    \centering
    \includegraphics[scale=0.6]{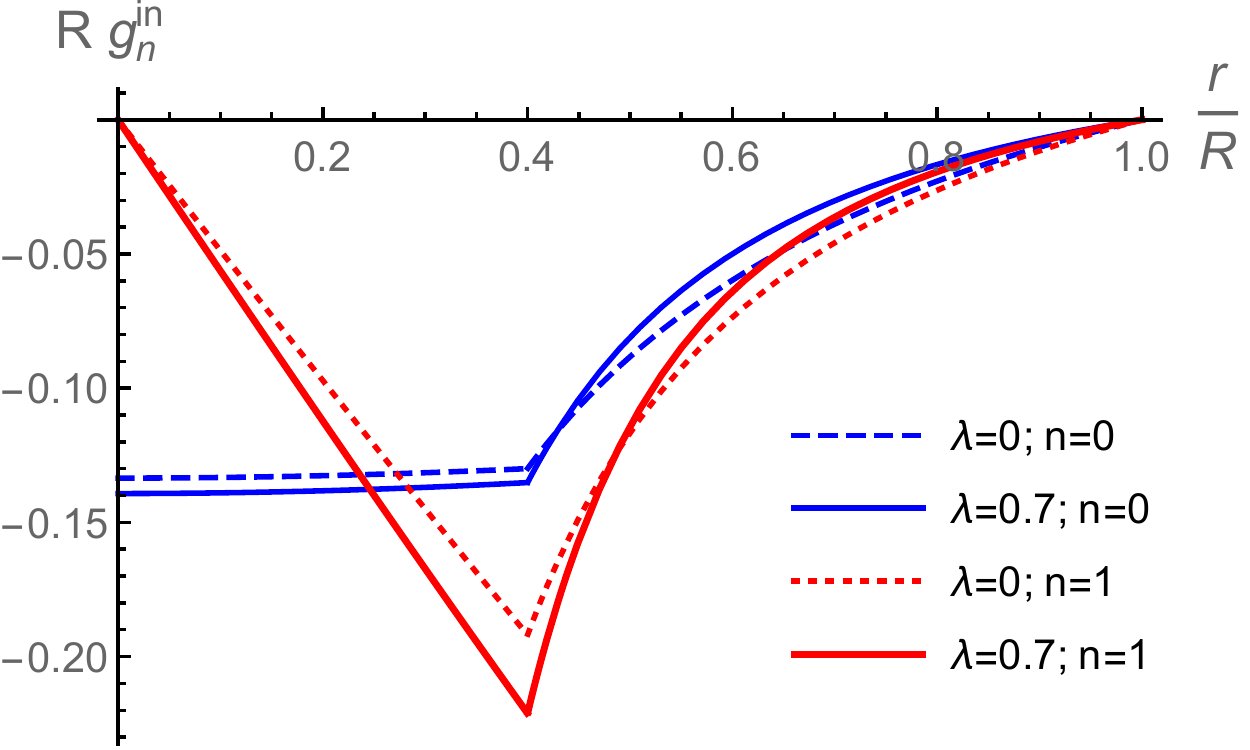}
    \caption{The reduced GF $R \, g _{n} ^{\mbox{\scriptsize in}}$ (\ref{Green}) for a massless scalar field (in a $3-$dimensional sphere) as a function of $r/R$ with the values $r ^{\prime} = 0.4\, R$ and $\omega R = 1$.} \label{gin}
\end{figure}

\begin{figure}[H]
    \centering
    \includegraphics[scale=0.6]{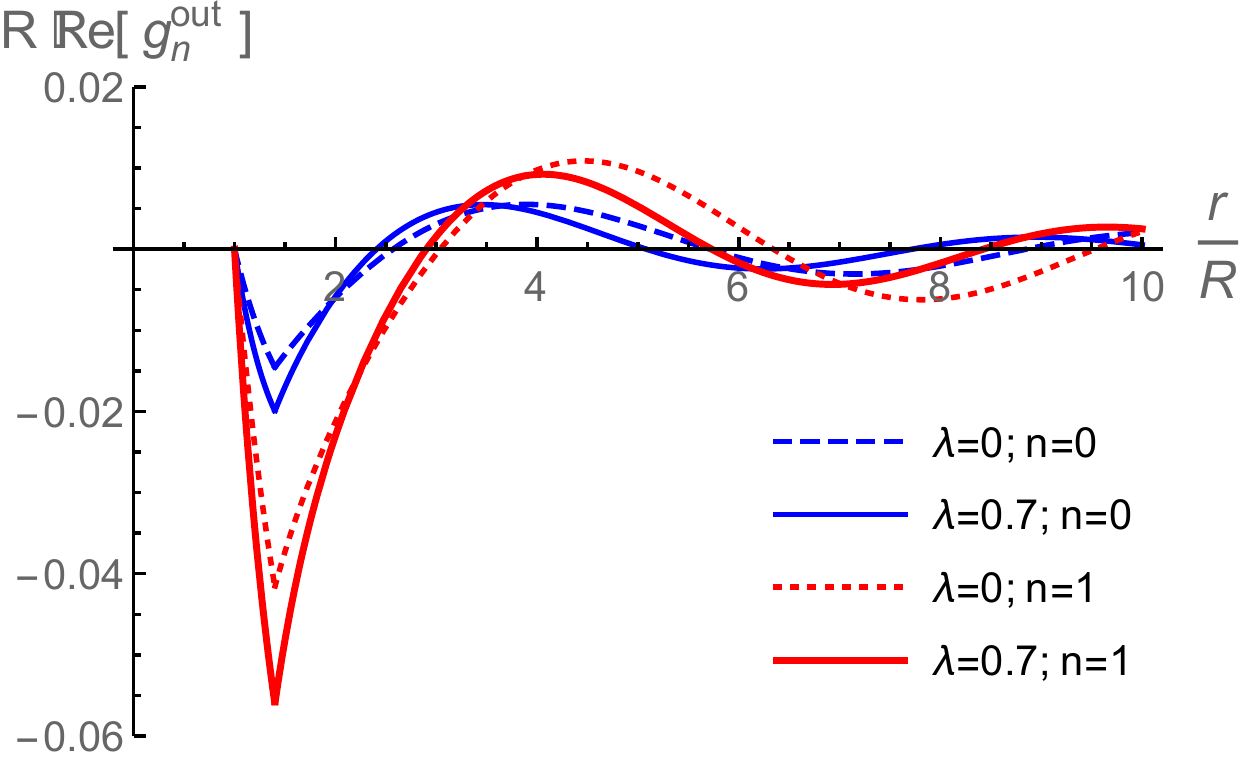}
    \caption{The real part of the reduced GF $ R\, g _{n} ^{\mbox{\scriptsize out}}$ (\ref{Green2}) for massless scalar field (in a $3-$dimensional sphere) as a function of $r/R$ with the values $r ^{\prime} = 1.4 R$ and $\omega R = 1$.}
    \label{gout}
\end{figure}

Next, making the change of variable $z=\cos\theta$ in Eq. (\ref{Angular eq}) we obtain
\begin{align}
\left[\, (1-z ^{2})  \frac{d ^{2}}{dz ^{2}} - z \,(D-1)\, \frac{d}{dz} + n(n+D-2) \,\right] \Theta(z) = 0 ,
\end{align}
which can be identified as the Gegenbauer equation \cite{Abramowitz}. Its regular solutions for $|z|=1$ are the ultraspherical or Gegenbauer polynomials
\begin{align}
\Theta _{n} (z) = C _{n} ^{(-1+D/2)} (z) , \qquad n \in \mathbb{Z} ^{+} \cup\{0\} \ . \label{Sap}
\end{align}
Therefore, the most general solution for (\ref{GFLV ur}) can be expressed as a linear superposition of the separated-variable solutions
\begin{align}
G _{\omega} (r,r ^{\prime} , z) = \sum _{n = 0} ^{\infty} g _{n} (r,r ^{\prime}) \,  C _{n} ^{(-1+D/2)} (z) \,  . \label{G1omega}
\end{align}
Now, we proceed to determine the GF in the inner and outer regions of the sphere. To this end, we recall that the reduced GF $g _{n} (r,r ^{\prime})$ must satisfy the Dirichlet BC (\ref{Dirichlet}) on the sphere, finiteness at the origin (\ref{Finiteness}), continuity at $r = r ^{\prime}$
\begin{align}
\lim _{\epsilon \rightarrow 0 ^{+}} \, g _{n} (r,r ^{\prime}) \big| ^{r = r ^{\prime} + \epsilon} _{ r = r ^{\prime} - \epsilon} = 0 \, ,
\end{align}
as well as the discontinuity in the first derivative
\begin{align}
\lim _{\epsilon \rightarrow 0 ^{+}} \, \frac{d g _{n} (r,r ^{\prime})}{dr} \bigg| _{r = r ^{\prime} -  \epsilon} ^{r = r ^{\prime} + \epsilon} = \frac{(2n + D - 2)}{4 \Lambda _{R} \pi ^{D/2} \, r ^{\prime \, D-1} } \Gamma \left( \frac{D-2}{2} \right) \,  ,
\end{align}
which follows from integration of Eq. (\ref{GFLV ur}) in the vicinity of $r ^{\prime}$.  Also, its behavior at $r \to \infty$ should decay appropriately. For the interior region of the sphere the general solution of Eq. (\ref{gnylv}) can be written as
\begin{align}
g _{n} ^{\mbox{\scriptsize in}} (r,r ^{\prime}) &= \ \left\lbrace \begin{array}{l} r ^{\alpha} a _{n} \, J _{s _{n}} (\Omega r) ,  \\[7pt] r ^{\alpha} \left[ b _{n} \, J _{s _{n}} (\Omega r) + c _{n} \, Y _{s _{n}} (\Omega r) \right] , \end{array} \begin{array}{c} r<r ^{\prime} < R \\[7pt] r ^{\prime} < r < R \end{array}  \right. \,, \label{gnin}
\end{align}
where $a _{n},b_{n}$ and $c_{n}$ are arbitrary parameters. In Eq. (\ref{gnin}), for $r<r ^{\prime} < R$, we have eliminated the linearly independent solution $r ^{\alpha} \,Y_{s _{n}} (\Omega r)$, which is singular at $r = 0$. This becomes clear when considering the asymptotic form of $Y _{s _{n}} (\Omega r)$ for small argument. In that case, the leading contribution in $r ^{\alpha} \,Y_{s _{n}} (\Omega r)$ is of the form $r ^{\alpha - s _{n}}$, which diverges at the origin provided that $s _{n} > \alpha$. The latter condition yields $D>2$ for all $n$, which is exactly the same condition obtained in the Lorentz symmetric case. This is why we stick to the case $D>2$. The special case $D=1$ will be considered separately. The BCs lead to the system of equations
\begin{align}
b _{n} \, J _{s _{n}} (\Omega R)\  + \ c _{n} \, Y _{s _{n}} (\Omega R) &\ = \ 0 , \notag \\ b _{n} \, J _{s _{n}} (\Omega r ^{\prime})\  + \ c _{n} \, Y _{s _{n}} (\Omega r ^{\prime})\ - \ a _{n} \, J _{s _{n}} (\Omega r ^{\prime}) &\ = \ 0 , \notag \\  b _{n} \, J _{s _{n}} ^{\prime} (\Omega r ^{\prime})\ + \ c _{n} \, Y _{s _{n}} ^{\prime} (\Omega r ^{\prime})\ -\  a _{n} \, J _{s _{n}} ^{\prime} (\Omega r ^{\prime}) &\ = \frac{(2n + D - 2)}{4 \Lambda_R \Omega \pi ^{D/2} r ^{\prime \, \alpha + D-1} } \Gamma \left( \frac{D-2}{2} \right)\ ,
\end{align}
which determine the coefficients $a _{n}$, $b _{n}$ and $c _{n}$. In that way, the reduced GF for the interior region takes the form
\begin{align}
 g _{n} ^{\mbox{\scriptsize in}}  (r,r ^{\prime}) \ =\  \frac{(2n + D - 2) \Gamma (-1+D/2) }{8 \Lambda_R \pi ^{\frac{D-2}{2}} }  \frac{r ^{\alpha}}{ r ^{\prime \, \alpha + D-2}} \frac{J _{s _{n}} (\Omega r _{<})}{J _{s _{n}} (\Omega R)} \left[ J _{s _{n}} (\Omega R)  Y _{s _{n}} (\Omega r _{>}) \,- \,J _{s _{n}} (\Omega r _{>}) Y _{s _{n}} (\Omega R) \right] , \label{Green}
\end{align}
where $r _{>}$ ($r _{<}$) is the greater (lesser) between $r$ and $r ^{\prime}$. In Fig. \ref{gin} we plot the reduced GF of Eq. (\ref{Green}) for a massless scalar field in 3 dimensions as a function of the dimensionless radius $r/R$ for $r ^{\prime} / R = 0.4$ and different values of $n$ and $\lambda$.
For the outer region, the general solution of Eq. (\ref{gnylv}) can be expressed in terms of the Hankel functions of the first and second kind:
\begin{align}
g _{n} ^{\mbox{\scriptsize out}}  (r,r ^{\prime}) =  \left\lbrace \begin{array}{l} r ^{\alpha} \left[ d _{n} \, H _{s _{n}} ^{(1)} (\Omega r) + e _{n} \, H _{s _{n}} ^{(2)} (\Omega r) \right] , \\[7pt] r ^{\alpha} f _{n} \, H _{s _{n}} ^{(1)} (\Omega r) , \end{array} \begin{array}{c} R < r<r ^{\prime} \\[7pt] R< r ^{\prime} < r \end{array}  \right. \, . \label{gnout}
\end{align}
The single Hankel function in the region $R<r ^{\prime} < r$ is required by the BC that $g _{n} ^{\mbox{\scriptsize out}}$ remains finite for $r \to \infty$ and has the asymptotic form of an outgoing plane wave. Again, the BCs lead to a set of equations for the coefficients $d _{n}$, $e _{n}$ and $f _{n}$, namely
\begin{align}
 d _{n} \, H _{s _{n}} ^{(1)} (\Omega R) + e _{n} \, H _{s _{n}} ^{(2)} (\Omega R)  &= 0 , \notag \\ f _{n} \, H _{s _{n}} ^{(1)} (\Omega r ^{\prime})  - d  _{n} \, H _{s _{n}} ^{(1)} (\Omega r ^{\prime}) - e _{n} \, H _{s _{n}} ^{(2)} (\Omega r ^{\prime}) &= 0 , \notag \\ f _{n} \, H _{s _{n}} ^{(1) \, \prime} (\Omega r ^{\prime})  - d  _{n} \, H _{s _{n}} ^{(1)  \, \prime} (\Omega r ^{\prime}) - e _{n} \, H _{s _{n}} ^{(2)  \, \prime} (\Omega r ^{\prime}) & = \frac{(2n + D - 2) \, \Gamma (-1 + D/2)}{4 \Omega  \Lambda_R \pi ^{D/2} r ^{\prime \, \alpha + D-1} } .
\end{align}
By solving the above system of equations, we arrive to the solution
\begin{align}
g _{n} ^{\mbox{\scriptsize out}}  (r,r ^{\prime}) (r,r ^{\prime}) = \frac{(2n + D - 2) \, \Gamma (-1 + D/2)}{16 i  \Lambda_R \pi ^{-1+ D/2} } \frac{r ^{\alpha}}{r ^{\prime \, \alpha + D - 2} }  \frac{H _{s _{n}} ^{(1)} (\Omega r _{>})}{H _{s _{n}} ^{(1)} (\Omega R)} \left[ H _{s _{n}} ^{(1)} (\Omega R) H _{s _{n}} ^{(2)} (\Omega r _{<}) - H _{s _{n}} ^{(1)} (\Omega r _{<})  H _{s _{n}} ^{(2)} (\Omega R) \right]\  , \label{Green2}
\end{align}
where, as before, $r _{>}$ ($r _{<}$) is the greater (lesser) between $r$ and $r ^{\prime}$.  In Fig. \ref{gout} we plot the real part of the reduced GF of Eq. (\ref{Green2}) for a massless scalar field in 3 dimensions as a function of the dimensionless radius $r/R$ for $r ^{\prime} / R = 1.4$ and different values of $n$ and $\lambda$. With the above GFs, we are in position to evaluate the Casimir force (\ref{FAbyG}) for the radial  spacelike case.

\subsection{Green's function: timelike case} \label{GFT1a2}

Our objective now is to derive the GF appearing in Eq. (\ref{GF eq omega}) for the timelike case $u ^{\mu} = (1,0,0,\cdots,0)$. In the polar variables ($r,\theta$) the function $G_{\omega} (r,r ^{\prime} , \theta)$ (\ref{GFT}) satisfies
\begin{align}
\bigg[ \Lambda _{T} \, \omega ^{2} - m ^{2} + \frac{\partial ^{2}}{\partial r ^{2}} + \frac{D-1}{r} \frac{\partial}{\partial r} + \frac{\sin ^{2-D} \theta}{r ^{2}} \frac{\partial}{\partial \theta} \left( \sin ^{D-2} \theta  \frac{\partial}{\partial \theta}  \right) \bigg] G _{\omega} (r,r ^{\prime} , \theta) = \frac{\delta (r-r ^{\prime}) \delta (\theta) \Gamma \left( \frac{D-1}{2} \right)}{2 \pi ^{\frac{D-1}{2}} r ^{D-1} \sin ^{D-2} \theta } , \label{GFLVUT}
\end{align}
where $\Lambda _{T} = 1 + \lambda$.  We observe that, unlike the radial spacelike case where Lorentz violation enters into the GF equation (\ref{GFLV ur}) in a nontrivial fashion, in the timelike case the LV-dependence can be absorbed into the frequency through the redefinition $\sqrt{\Lambda _{T}} \, \omega \rightarrow \omega$, thus leaving us with the Lorentz-symmetric GF equation. Therefore, we safely take for granted the reduced GFs for the timelike case. Since the spherical symmetry is untouched in both cases, the angular part is still given by Eq. (\ref{Sap}). However, for the inner region the reduced GF $\tilde g _{n} (r,r ^{\prime})$ is
\begin{align}
 \tilde{g} _{n} ^{\mbox{\scriptsize in}} (r,r ^{\prime}) = \frac{(2n + D - 2) \Gamma (-1+D/2) }{8 \pi ^{\frac{D-2}{2}} }  \frac{1}{ (r\, r ^{\prime} )^{ \, -1+\frac{D}{2}} } \frac{J _{\nu} (\Omega ^{\prime} r _{<})}{J _{\nu} (\Omega ^{\prime} R)} \left[ J _{\nu} (\Omega ^{\prime} R)  Y _{\nu} (\Omega ^{\prime} r _{>}) - J _{\nu} (\Omega ^{\prime} r _{>}) Y _{\nu} (\Omega ^{\prime} R) \right] ,
\label{GreenIn}
\end{align}
while for the outer region we obtain the solution
\begin{align}
\tilde{g} _{n} ^{\mbox{\scriptsize out}} (r,r ^{\prime}) = \frac{(2n + D - 2) \, \Gamma (-1 + D/2)}{16 i  \pi ^{-1+ D/2} } \frac{1}{ (r \, r ^{\prime} )^{ \, -1+\frac{D}{2}} }  \frac{H _{\nu} ^{(1)} (\Omega ^{\prime} r _{>})}{H _{\nu} ^{(1)} (\Omega ^{\prime} R)} \left[ H _{\nu} ^{(1)} (\Omega ^{\prime} R) H _{\nu} ^{(2)} (\Omega ^{\prime} r _{<}) - H _{\nu} ^{(1)} (\Omega ^{\prime} r _{<})  H _{\nu} ^{(2)} (\Omega ^{\prime} R) \right] , \label{Green2In}
\end{align}
where $\nu = n-1+D/2$, $\Omega ^{\prime} = \sqrt{\Lambda_{T} \, \omega ^{2} - m ^{2}}$, and $r _{>}$ ($r _{<}$) is the greater (lesser) between $r$ and $r ^{\prime}$. As we shall see later, the fact that in the present case the parameter $\Lambda _{T}$ for Lorentz violation enters in a simple manner into the GF will be reflected in the expression for the Casimir pressure as a global factor.

\section{Casimir effect}\label{CEsection}

In this section we present for the cases I and II the formal expression for the Casimir force (\ref{FAbyG}), with $D>2$, explicitly. Substituting the Fourier representation of the GF (\ref{GFT-inv})  into Eq. (\ref{FAbyG}) we obtain an expression for the Casimir pressure in terms of the frequency-dependent GF:
\begin{align}
\frac{F}{A} = - \frac{i}{2} \, \Lambda \int _{- \infty} ^{\infty} \, \frac{d \omega}{2 \pi} \, \lim _{\vec{x} ^{\, \prime} \rightarrow \vec{x}} \bigg[ \frac{\partial }{\partial r}\frac{\partial}{\partial r ^{\prime}} G _{\omega}^{(\mbox{\scriptsize in})}( \vec{x} , \vec{x} ^{\, \prime} ) - \frac{\partial }{\partial r} \frac{\partial}{\partial r ^{\prime}} G _{\omega}^{(\mbox{\scriptsize out})}( \vec{x} , \vec{x} ^{\, \prime} )  \bigg] \bigg\vert _{\| \vec{x}\|\,=\,R}    .
\end{align}
This can be further simplified by using the expansion (\ref{G1omega}). Upon substitution we find
\begin{align}
\frac{F}{A} = - \frac{i}{4 \pi} \, \Lambda \sum _{n = 0} ^{\infty} \frac{\Gamma (n +D - 2 )}{n! \, \Gamma ( D - 2 )} \int _{- \infty} ^{\infty} \, d\omega \, \lim _{r ^{\prime} \rightarrow r} \bigg[ \frac{\partial }{\partial r}\frac{\partial}{\partial r ^{\prime}} g _{n} ^{\mbox{\scriptsize in}} (r,r ^{\prime}) - \frac{\partial }{\partial r} \frac{\partial}{\partial r ^{\prime}} g _{n} ^{\mbox{\scriptsize out}} (r,r ^{\prime})  \bigg] \bigg\vert _{r = R}   , \label{CasPressureGeneral}
\end{align}
where we have used the value of the ultraspherical polynomials at $z = 1$ (which corresponds to $\theta = 0$),
\begin{align}
C _{n} ^{( \alpha )} (1) =   \frac{\Gamma (n+ 2 \alpha )}{n! \, \Gamma (2 \alpha )} .
\end{align}
Having determined the above general expression for the Casimir pressure in terms of the reduced GF, we can now evaluate it explicitly for each case.

\subsection{Case I: radial spacelike case}

From the reduced GFs $g ^{\mbox{\scriptsize in/out}} (r,r ^{\prime})$, given by Eqs. (\ref{Green}) and (\ref{Green2}), one can directly compute the limit appearing in the integrand of Eq. (\ref{CasPressureGeneral}). For the interior region, without loss of generality we take $r _{<} = r ^{\prime}$ and $r _{>} = r$ in Eq. (\ref{Green}), wherefrom we obtain
\begin{align}
\lim _{r ^{\prime} \rightarrow r} \frac{\partial }{\partial r}\frac{\partial}{\partial r ^{\prime}} g _{n} ^{\mbox{\scriptsize in}} (r,r ^{\prime}) \bigg\vert _{r = R}  = \frac{(2n + D - 2) \Gamma (-1+D/2) }{4 \Lambda_R \pi ^{D/2 } R ^{D}}   \left[ \Omega R  \frac{J _{s _{n}} ^{\prime} (\Omega R)}{J _{s _{n}} (\Omega R)} - (\alpha + D-2) \right] , \label{partialrr}
\end{align}
where we have used that the Wronskian of $J _{\nu} (z)$ and $Y _{\nu} (z)$ has the value $W [ J _{\nu} (z) , Y _{\nu} (z) ] = 2/ ( \pi z )$, and the prime in the Bessel function denotes derivative with respect to its argument. The converse choice between $r _{>}$ and $r _{<}$ produces, at the end of the calculations, the same final result for the Casimir force. A similar procedure for the exterior region (for which we take $r _{>} = r ^{\prime}$ and $r _{<} = r$) yields
\begin{align}
\lim _{r ^{\prime} \rightarrow r} \frac{\partial }{\partial r}\frac{\partial}{\partial r ^{\prime}} g _{n} ^{\mbox{\scriptsize out}} (r,r ^{\prime}) \bigg\vert _{r = R} = -  \frac{(2n + D - 2) \, \Gamma (-1 + D/2)}{4 \Lambda_R \pi ^{ D/2} R ^{D} }  \left[ \Omega R \frac{H _{s _{n}} ^{(1)  \, \prime} (\Omega R)}{H _{s _{n}} ^{(1)} ( \Omega R)} - (\alpha + D - 2) \right] ,
\end{align}
where we have employed that the Wronskian of $H _{\nu} ^{(1)} (z)$ and $H _{\nu} ^{(2)} (z)$ is $W [ H _{\nu} ^{(1)} (z) , H _{\nu} ^{(2)} (z) ] = 4 / (i \pi z)$. Finally, inserting these results into Eq. (\ref{CasPressureGeneral}) we obtain the following expression for the Casimir force
\begin{align}
\frac{F_r}{A} = - \sum _{n = 0} ^{\infty} i \,\frac{(n -  1+ D/ 2) \,\Gamma (n+D-2)}{ (2R) ^{D} \pi ^{\frac{D+1}{2}} n ! \, \Gamma \left( \frac{D-1}{2} \right)}   \int _{- \infty} ^{\infty} d \omega  \left[ \Omega R \frac{J _{s _{n}} ^{\prime} (\Omega R)}{J _{s _{n}} (\Omega R)} + \Omega R \frac{H _{s _{n}} ^{(1)  \, \prime} (\Omega R)}{H _{s _{n}} ^{(1)} ( \Omega R)} - 2 (\alpha + D - 2)  \right]   .
\end{align}
where we have used the duplication formula $\Gamma (2 \alpha ) = 2 ^{2 \alpha -1} \Gamma (\alpha ) \Gamma (\alpha + 1/2) / \sqrt{\pi}$. For the massless case $m = 0$ ($\Omega = \omega / \sqrt{\Lambda _{R}}$) the rotation of $\pi /2$ in the complex-$\omega$ plane, namely $x = i \, \omega\, R / \sqrt{\Lambda _{R}}$, transforms the above expression into
\begin{align}
\frac{F_r}{A} = - \sqrt{ \Lambda_R} \sum _{n = 0} ^{\infty} \frac{(n -  1+ D/ 2) \Gamma (n+D-2)}{2 ^{D-1} R ^{D+1} \pi ^{\frac{D+1}{2}} n ! \, \Gamma \left( \frac{D-1}{2} \right)}   \int _{0} ^{\infty} d x \left[ x \frac{I _{s _{n}} ^{\prime} (x)}{I _{s _{n}} (x)} + x \frac{K _{s _{n}} ^{\prime} (x)}{K _{s _{n}} ( x)} - 2 (\alpha + D - 2)  \right] . \label{FLVR}
\end{align}
In the limit $\lambda \to 0$ we obtain $s _{n} \to n-1+\frac{D}{2}$ and $\alpha\rightarrow 1-D/2$, which corresponds to the Lorentz-invariant result provided that $D>2$ \cite{Bender-Milton}. Hereafter, we will restrict ourselves to the massless case $m=0$.

\subsubsection*{Special case $D=1$ }

Formally, Eq. (\ref{FLVR}) was derived under the assumption $D>2$. However, in the particular case $D=1$ the expression (\ref{FLVR}) is well defined and can be evaluated analytically. At $D=1$, the series appearing in Eq. (\ref{FLVR}) truncates after two terms. This happens because of the identity
\begin{align}
\lim _{D \rightarrow 1} \frac{\Gamma(n+D-2)}{\Gamma(\frac{D-1}{2})}\ = \ -\frac{1}{2}\delta_{n0}\  + \ \frac{1}{2}\delta_{n1} \ . \label{identi1}
\end{align}
Hence, only the terms with $n=0$ and $n=1$ will give contribution. As mentioned in Sec. \ref{GFRa}, this implies that the Eq. (\ref{radial1a}), up to a redefinition of the frequency $\omega\rightarrow \omega\,\sqrt{\Lambda_R}$ (equivalently, $x \rightarrow x \,\sqrt{\Lambda_R}$), can be cast in the form of the Lorentz invariant case for which $\alpha\to1/2$ and the Bessel function order is $\nu \to n-1/2$, see Ref. \cite{Bender-Milton}. Therefore, the Casimir force in the presence of Lorentz violation is given by
\begin{align}
\frac{F_{r} ^{(D=1)}}{A} &=  \sqrt{\Lambda_R}\,\times \,\frac{-1}{4 \pi R ^{2}}  \int _{0} ^{ \infty} d y \left[ y \frac{I _{-\frac{1}{2}} ^{\prime} (y)}{I _{-\frac{1}{2}} (y)}+y \frac{I _{\frac{1}{2}} ^{\prime} (y)}{I _{\frac{1}{2}} (y)}  + y \frac{K _{-\frac{1}{2}} ^{\prime} (y)}{K _{-\frac{1}{2}} (y)}+ y \frac{K _{\frac{1}{2}} ^{\prime} (y)}{K _{\frac{1}{2}} (y)} +2 \right]\, . \label{F2R}
\end{align}
where the second factor is just the Lorentz invariant result derived in \cite{Bender-Milton}. We emphasize that this situation occurs only for $D=1$. Now, using the explicit forms of the Bessel functions required by Eq. (\ref{F2R}) and evaluating the integral over $y$ we get
\begin{align}
\frac{F _{r} ^{(D=1)}}{A} &= - \sqrt{\Lambda_R}\,\frac{\pi}{96\, R^2} .  \label{FR1a}
\end{align}
Thus, in the case $D=1$, the Casimir stress (\ref{FR1a}) can be greater or smaller than the Lorentz-symmetric case depending on the sign of $\lambda$. In particular, recalling that $\vert \lambda \vert < 1$, $\lambda < 0$ tends to increase the force, while $\lambda > 0$ produces the opposite effect.

\subsection{Case II: timelike case}

In this case we have to evaluate Eq. (\ref{CasPressureGeneral}) with the reduced GFs $\tilde{g} ^{\mbox{\scriptsize in/out}} (r,r ^{\prime})$ of Eqs. (\ref{GreenIn}) and (\ref{Green2In}). As discussed in Sec. \ref{GFT1a2}, the timelike case behaves exactly as the Lorentz symmetric case with a simple redefinition of the frequency. Taking similar steps as in the previous section, after some algebra we find that for the timelike case the Casimir force takes the form
\begin{align}
\frac{F_{t}}{A} = - \frac{1}{\sqrt{ \Lambda} _{T}} \sum _{n = 0} ^{\infty} \frac{(n -  1+ D/ 2) \Gamma (n+D-2)}{2 ^{D-1} R ^{D+1} \pi ^{\frac{D+1}{2}} n ! \, \Gamma \left( \frac{D-1}{2} \right)}   \int _{0} ^{\infty} d x \left[ x \frac{I _{\nu} ^{\prime} (x)}{I _{\nu} (x)} + x \frac{K _{\nu} ^{\prime} (x)}{K _{\nu} ( x)} + 2 -D  \right] , \label{FLVT}
\end{align}
where $\nu = n - 1 + D/2$. Clearly, this expression is proportional to that obtained in the absence of Lorentz violation ($\lambda=0, \Lambda_T=1$), i.e.
\begin{align}
\frac{F _{t} (\Lambda _{T})}{A} \ = \ \frac{1}{\sqrt{ \Lambda} _{T}} \, \frac{F _{t} (\Lambda_{T} = 1)}{A}  . \label{FTR1}
\end{align}
Therefore, if $\lambda < 0$ the Casimir stress (\ref{FTR1}) is enlarged as compared with that of the Lorentz-symmetric case whilst it is diminished when $\lambda > 0$. In particular, for the one-dimensional case $D=1$ the Eq. (\ref{FTR1}) reduces to
\begin{align}
\frac{F_ {t} ^{(D=1)}}{A} & \ = \ - \frac{1}{\sqrt{\Lambda _{T}}}\,\frac{\pi}{96\, R ^{2}}\ ,
\end{align}
c.f. (\ref{FR1a}). We emphasize that, unlike the radial spacelike case, the factorization in (\ref{FTR1}) holds for an arbitrary dimension $D$. The LV contributions appear in the global multiplicative factor ${ \Lambda} _{T}^{-1/2}$ only.

\section{Towards numerical Evaluation of the Casimir force}\label{Numerical Section}

The infinite series in Eqs. (\ref{FLVR}) and (\ref{FLVT}) are divergent. They do not even exist for some values of $D$ where the poles of the Gamma function take place. Partly, this behavior is a manifestation of the nonzero vacuum energy inherent to the CE. Therefore, in order to remove non-physical divergences a method of regularization is required. In this Section, we adapt to the problem at hand the regularization method used in Ref. \cite{Bender-Milton}.

\subsection{Radial spacelike case}

In this section we manipulate the form of the expression (\ref{FLVR}) so that each integral in the series exists. Afterwards, we can evaluate the resulting series numerically. To render the integrals finite we should remove the contact terms that arise from the local behavior near the boundaries. This can be done by replacing the constant $- 2(\alpha + D - 2)$ in Eq. (\ref{FLVR}) by $1$, leaving unchanged the value of the integral (\ref{FLVR}), and yet leads to well-defined (finite) integrals. For the formal proof see Refs. \cite{Bender-Milton,Milton's book,Milton 3} . Eventually, after integration by parts, the series (\ref{FLVR}) can be rewritten in the form
\begin{align}
\frac{F_{r}}{A} = \sqrt{ \Lambda_R} \, \sum _{n = 0} ^{\infty} \frac{(n -  1+ D/ 2) \Gamma (n+D-2)}{2 ^{D-1} R ^{D+1} \pi ^{\frac{D+1}{2}} n ! \, \Gamma \left( \frac{D-1}{2} \right)}   \int _{0} ^{\infty} d x \,\textrm{ln}\left[ 2x I_{s_n}(x)K_{s_n}(x) \right] . \label{Rd2}
\end{align}
Now, at a noneven integer $D>2$ each term of the above series exists for any value of $n$. However, the above expression is still not useful since the series does not converge. In order to obtain a convergent reformulation, we analyze the asymptotic behavior of the integrals in Eq. (\ref{Rd2}) for large $n$. This will help us to better understand the source of divergences. From the uniform asymptotic expansions of the modified Bessel functions for large \mbox{$\mu$ \cite{Abramowitz}},
\begin{align}
I _{\mu} (\mu z) \sim \frac{e^{\mu \eta}}{\sqrt{2\pi \mu}(1+z^2)^{\frac{1}{4}}}\sum_{k=0} ^{\infty} \frac{U _{k} (p)}{ \mu ^{k}} , \quad\quad\quad
K _{\mu} (\mu z) \sim \bigg( \frac{\pi}{2\mu} \bigg) ^{\frac{1}{2}} \frac{e ^{- \mu \eta}}{(1+z^2)^{\frac{1}{4}}} \sum _{k=0} ^{\infty} (-1) ^{k} \frac{U _{k} (p)}{\mu ^{k}} ,
\label{expansiones}
\end{align}
where
\begin{align}
\eta = (1+z^2) ^{\frac{1}{2}} + \textrm{ln} \frac{z}{1+\sqrt{1+z^2}} \quad , \qquad \quad p = (1+z^2)^{-\frac{1}{2}}
\end{align}
and
\begin{align}
U _{0} (p) &= 1 , \notag \\ U _{1} (p) &= \frac{1}{24}(3 p - 5 p ^{3}) , \notag \\ U _{2} (p) &= \frac{1}{1152}(81 p ^{2} - 462 p ^{4} + 385 p ^{6}) , \notag \\ U _{3} (p) &= \frac{1}{414720}(30375 p ^{3} - 369603 p ^{5} + 765765 p ^{7} - 425425 p ^{9}),
\end{align}
we obtain the asymptotic expansion of integrals
\begin{align}
Q _{n} & \equiv - \int _{0} ^{\infty} dx \, \textrm{ln}[2 x I _{s _{n}}(x) K _{s _{n}}(x)] = - s _{n} \int _{0} ^{\infty} dy \, \textrm{ln}[2 s _{n} y I _{s _{n}}(s _{n} y) K _{s _{n}}( s _{n} y)] \notag \\ & \phantom{=} \sim \frac{\nu\pi}{2\sqrt{\Lambda_R}} + \frac{\pi \sqrt{\Lambda _{R}}}{128 \nu} -\frac{35 \pi (\sqrt{\Lambda _{R}}) ^{3}}{32768 \nu ^{3}} + \frac{565 \pi( \sqrt{\Lambda _{R}}) ^{5}}{1048576 \nu ^{5}} ,
\end{align}
for $n \rightarrow \infty$, where $s _{n} \sim \frac{\nu}{ \sqrt{\Lambda _{R}}}$ with $\nu=n-1+\frac{D}{2}$. The first term will give rise to divergences in Eq. (\ref{FLVR}), except for the special case $D=1$, where the series truncates. To solve this problem, let us introduce an analytic summation procedure based on the properties of the Riemann $\zeta$ function \cite{Bender-Milton}. The leading $n$-large behavior of the summand in Eq. (\ref{Rd2}) is given by
\begin{align}
- \frac{(n-1+\frac{D}{2})\Gamma(n+D-2)}{2^{D-1}\pi^{\frac{D+1}{2}}R^{D+1}n!\Gamma(\frac{D-1}{2})} Q _{n} \sim \frac{1}{2^{D} \pi ^{\frac{D+1}{2}} R ^{D+1}\Gamma(\frac{D-1}{2})} (c _{1} n ^{D-1} + c _{2} n ^{D-2} + \cdots + c _{k} n ^{D-k} + \cdots) , \quad (n \rightarrow \infty) , \label{Exp1}
\end{align}
where the coefficients $c _{k}$ depend on the dimension $D$ and the LV parameter $\Lambda _{R}$, i.e. $c _{k} = c _{k} (\Lambda _{R} , D)$. The first lowest coefficients are presented in Appendix \ref{AP1}. Since
\begin{align}
\sum _{n=1} ^{\infty} c _{k} n ^{D-k} = c _{k} \zeta(k-D) , \label{zeta1}
\end{align}
we can add to Eq. (\ref{Rd2}) $K$ terms of the form (\ref{zeta1}) and correspondingly subtract the right-hand side of Eq. (\ref{Exp1}) with the same number of terms. Thereby, the series (\ref{Rd2}) takes the form
\begin{align}
\frac{F _{r}}{A} = \sqrt{\Lambda _{R}} \bigg[ \frac{\Gamma(\frac{D}{2})}{4R^{D+1} \pi ^{1+\frac{D}{2}}} Q _{0} + \sum _{n=1} ^{\infty} \bigg( \frac{(n-1+\frac{D}{2})\Gamma(n+D-2)}{2 ^{D-1} \pi ^{\frac{D+1}{2}} R ^{D+1} n! \Gamma(\frac{D-1}{2})} Q _{n} + \frac{1}{2 ^{D} \pi ^{\frac{D-1}{2}} R ^{D+1}\Gamma(\frac{D-1}{2})} \sum _{k=1} ^{K} c _{k} n ^{D-k} \bigg) \nonumber  \\
-  \frac{1}{2 ^{D} \pi ^{\frac{D-1}{2}} R ^{D+1}} \sum _{k=1} ^{K} c _{k} \zeta(k-D)\bigg] , \label{RFLV}
\end{align}
and quickly converges since the $n$th term in the series tends to zero as $n ^{D-K-1}$, provided that $D<K$. Note that the Casimir stress is singular at all even positive integer values of $D$, where  Eq. (\ref{RFLV}) has simple poles at $D = 2N$, with $N = 1, 2, 3,\ldots $ .

\subsection{Timelike case}

For the timelike case we simply present the final result for the Casimir force given by
\begin{align}
\frac{F_{t}}{A} = \frac{1}{\sqrt{\Lambda_T}} \bigg[ \frac{\Gamma(\frac{D}{2})}{4\,R^{D+1} \pi ^{1+\frac{D}{2}}} Q _{0} + \sum_{n=1} ^{\infty} \bigg( \frac{(n-1+\frac{D}{2})\Gamma(n+D-2)}{2^{D-1} \pi ^{\frac{D+1}{2}} R ^{D+1} n! \Gamma(\frac{D-1}{2})} Q _{n} + \frac{1}{2 ^{D} \pi ^{\frac{D-1}{2}}R^{D+1}\Gamma(\frac{D-1}{2})} \sum_{k=1} ^{K} b _{k} n ^{D-k} \bigg) \nonumber \\
- \frac{1}{2^D \pi^{\frac{D-1}{2}}R^{D+1}} \sum_{k=1}^K b_k \zeta(k-D)\bigg] , \label{Ftr}
\end{align}
where the coefficients $b _{k}$ are those given in the Appendix \ref{AP1} evaluated at $\Lambda _{R} = 1$, i.e. $b _{k} = c _{k} \vert _{\Lambda _{R} = 1}$.

\section{Numerical results}\label{NumResultsSec}

We have evaluated numerically the expressions (\ref{RFLV}) and (\ref{Ftr}) for the radial spacelike and timelike cases, respectively. In particular, for $D=3$ and $\Lambda _{R} = 1$ (i.e. without LV) the Casimir stress we obtain is
\begin{align}
F _{r} ^{(D=3)} = \frac{0.0028172}{R ^{2}} , \label{Fr3dn}
\end{align}
which, up to the numerical precision used in the present study, agrees with the value $0.0028168 / R ^{2}$ reported by Bender and Milton in Ref. \cite{Bender-Milton}. We observe that, unlike the attractive Casimir stress between parallel plates, in this case the force is repulsive, i.e. it tends to inflate the sphere.

\begin{figure*}
    \centering
    \includegraphics[scale=0.45]{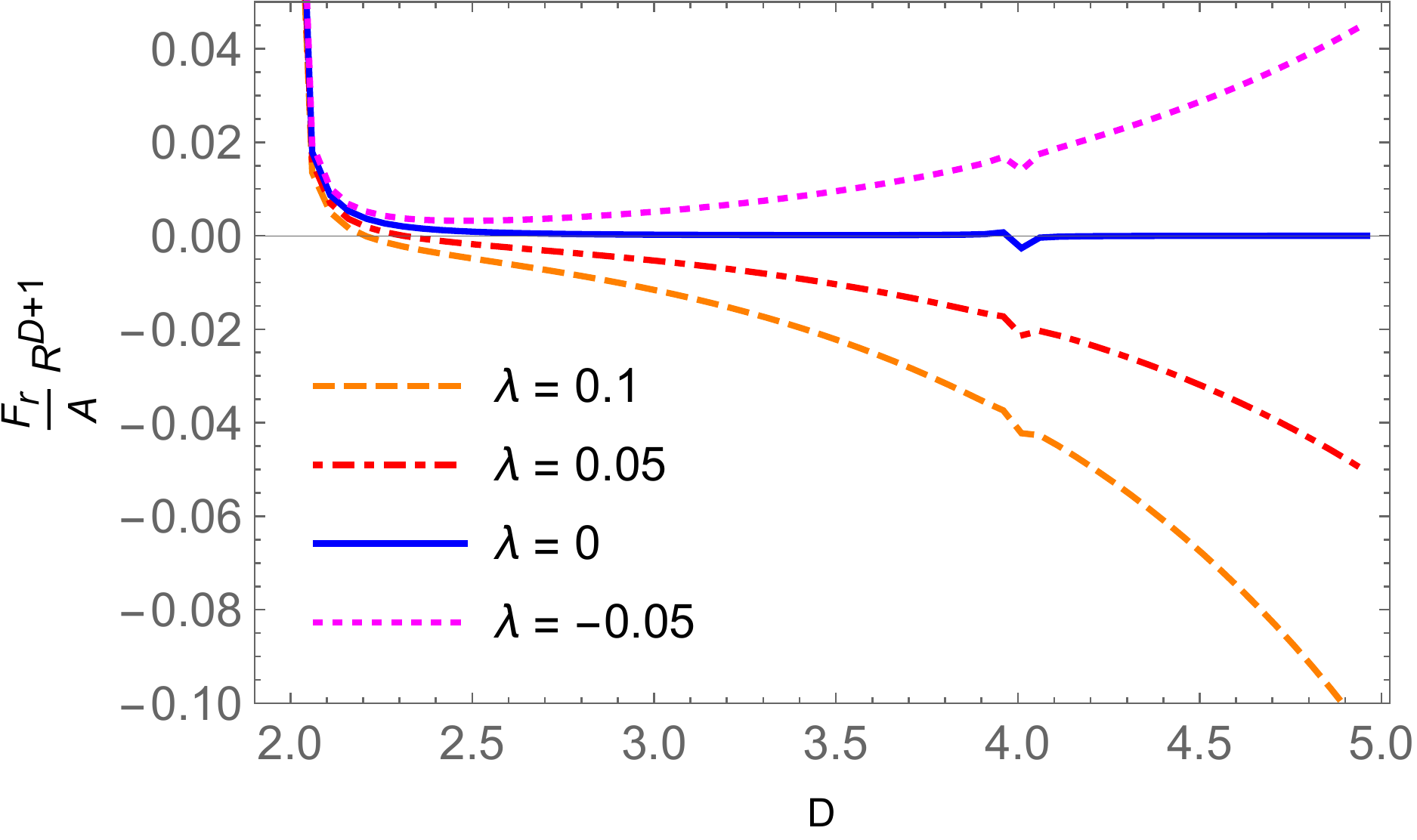}
    \caption{Plot of the Casimir pressure $F_{r}/A$ for a $D$-dimensional spherical shell, with $2<D<5$.}
    \label{ForceDR}
\end{figure*}

To visualize the LV contribution in more detail, in Fig. \ref{ForceDR} we plot the Casimir pressure $F_{r} / A$ as a function of the dimension $D$ for different values of the LV parameter $\lambda$. There we observe that the force can be repulsive or attractive, depending on the sign and strength of the parameter $\lambda$. Remarkably, we observe that for any $D>2$ there exists a \emph{critical value} $\lambda_c\neq0$ at which the Casimir force (\ref{RFLV}) vanishes, i.e.
\begin{align}
F _{r} (\lambda = \lambda _{c} ) = 0 . \label{ZeroCond}
\end{align}
The force $F_r$ is positive when $\lambda < \lambda _{c}$, and flips its sign for $\lambda > \lambda _{c}$ thus tending to implode the sphere. Such a transition does not occur in the Lorentz-symmetric case \cite{Bender-Milton} as well as in the LV scalar Casimir stress for parallel plates \cite{Escobar-Medel-Martin, PLB}.

\begin{figure*}
    \centering
    \includegraphics[scale=0.45]{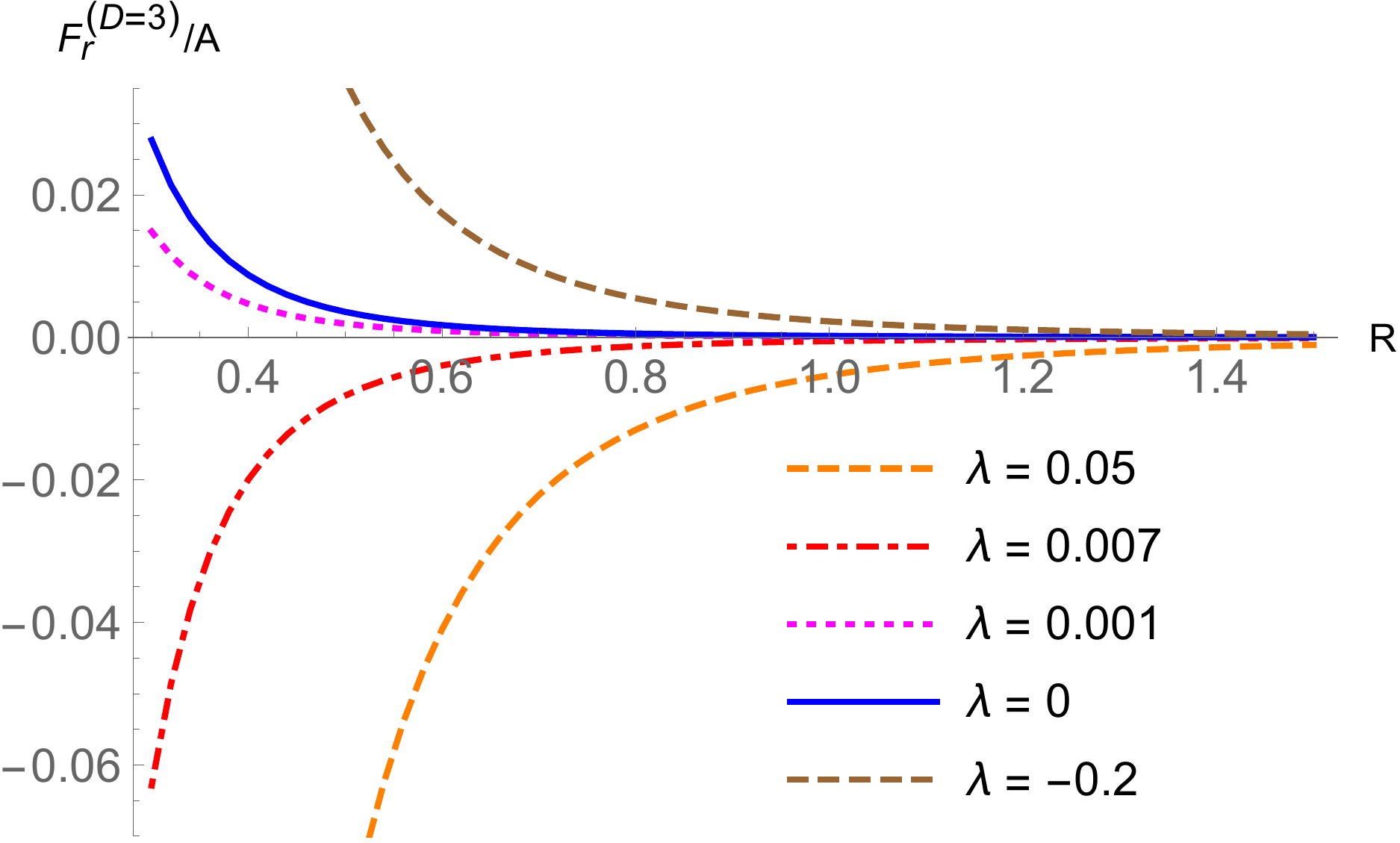}
    \includegraphics[scale=0.43]{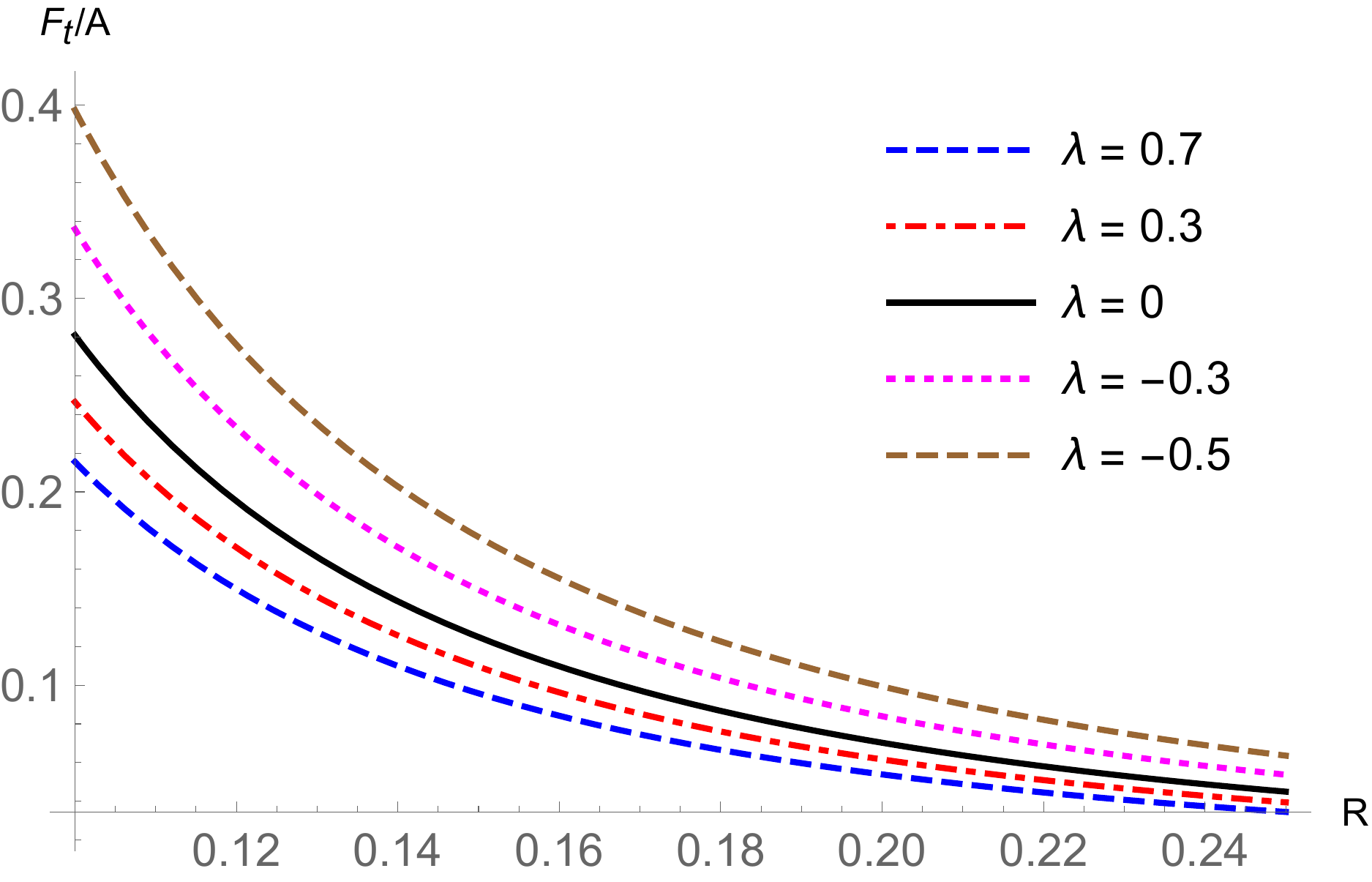}
       \caption{Plots of the Casimir pressure $F_{r}/A$ (left) and $F_{t}/A$ (right) for a massless three-dimensional scalar field as a function of the radius $R$.}
    \label{ForceD3R}
\end{figure*}

The most relevant physical case corresponds to $D=3$. In Fig. \ref{ForceD3R} we plot the Casimir pressure for the specific cases considered in this work as a function of $R$ and different values of $\lambda$. In the timelike case, presented in the right panel of Fig. \ref{ForceD3R}, the Casimir pressure in the presence of Lorentz violation behaves exactly as the usual attractive Lorentz-symmetric Casimir pressure. As Eq. (\ref{Ftr}) shows, the only difference is that the strength of $F _{t} /A$ can be either larger or smaller than $F _{t} (\Lambda _{t} = 1) /A$, depending on the sign of $\lambda$: a positive $\lambda$ yields to a smaller pressure, while a negative $\lambda$ renders to a larger pressure. On the other hand, as shown in the left panel of Fig. \ref{ForceD3R}, the radial spacelike case exhibits a more interesting behavior. In this case, the solution of Eq. (\ref{ZeroCond}) produces the critical value
\begin{align}
\lambda _{c} \approx 0.0025 . \label{CritLambda}
\end{align}
As we can see in the plot, the Casimir pressure changes sign when $\lambda$ transits across the critical value $\lambda _{c}$. From a high-energy physics perspective, the critical value of Eq. (\ref{CritLambda}) would not be admissible, since it is expected to be much smaller than one. However, Lorentz-violating effective field theories also emerge in condensed matter systems, where the symmetry breaking parameters are not necessarily small. For instance, topological insulators, which are bulk insulators with Dirac fermions on the surface \cite{TIs}, have enabled the theoretical possibility of realizing axion electrodynamics \cite{Wilczek} in a condensed matter system. Also, a Lorentz-violating extension of quantum electrodynamics can be realized with a novel class of materials known as Weyl semi-metals, which are systems which host low-energy quasiparticles that are described by the Weyl equations \cite{Grushin}. Undoubtedly, these kind of low-energy systems, where high-energy phenomena take place, represent a promising arena in which our predictions could be tested. An outstanding example of this is the chiral magnetic effect, which is an electric current along an externally applied magnetic field due to chirality imbalance. This effect was first predicted to occur in quark-gluon plasma, but it was experimentally observed in the Dirac semimetal ZrTe$_{5}$ \cite{CME}.

Reversing the Casimir force have been a topic of interest since its inception. To revert the sign, one must usually search non symmetric situations or vacuum mediated proposals. The first Casimir repulsion proposal, known as Dzyaloshinskii repulsion \cite{Dzyaloshinskii}, involves the presence of a dielectric fluid filling the space between two dielectrics in a parallel configuration. Recently, it was proposed that switching between repulsive and attractive Casimir forces could be realized with two topological insulator plates \cite{Cortijo, MCU}, as well as between two Weyl semimetallic plates \cite{Wilson}. In the context of the Lorentz-violating scalar field theory described by the Lagrangian (\ref{Lagrangian1}), it was recently shown that the Casimir force retains its attractive character in the parallel plate configuration \cite{Escobar-Medel-Martin, PLB}. However, as our results support, the Lorentz-violating parameter $\lambda$ allows us to tune between the standard repulsive to attractive Casimir force acting on a $D$-dimensional sphere, with $D>2$, when $u ^{\mu}$ points along the radial direction.

\section{Conclusions} \label{ConcluSection}

We have analyzed the effects of Lorentz symmetry violation in the scalar Casimir self-stress on a $D$-dimensional spherical shell with $D>2$. By considering a flat background spacetime, we have chosen two possible scenarios of violation of Lorentz symmetry through a fixed timelike $u ^{\mu} = (1,0,...,0)$ and a fixed spacelike $u ^{\mu} = (0,1,0,...,0)$ vectors. For each case the Casimir stress was obtained by using Green's function techniques.

In the timelike case, for any $D$ we found that Lorentz violation manifests in the Casimir force through a global rescaling factor only, i.e. $F _{t} (\Lambda _{T}) = F _{t} (\Lambda _{T} = 1) / \sqrt{\Lambda _{T}}$, where $\Lambda _{T} = 1 + \lambda$. This is because the Lorentz breaking term in the equation of motion (\ref{EQLV}) can be absorbed by redefining the frequency as $\omega \to \sqrt{\Lambda _{T}} \, \omega$, thus going back to the standard Klein-Gordon equation.

The radial spacelike case is quite more interesting. The corresponding GF depends on the parameter $\Lambda _{R} = 1 - \lambda$ in a nontrivial fashion. The expression for the Casimir pressure is not proportional to the Lorentz-symmetric case. The main difference with the timelike case lies on the $\lambda$-dependence in the order of the Bessel functions appearing in Eq. (\ref{RFLV}), thus making this case counter intuitive. The special case $D=1$ admits an analytical solution and it was discussed in detail. For dimensions $2\leq D\leq5$, we numerically evaluated the Casimir stress. An interesting aspect is that there exists a critical value $\lambda _{c}$ for which the Casimir stress vanishes. Moreover, for $\lambda < \lambda _{c}$ the Casimir stress is positive (tends to expand the sphere) while for $\lambda > \lambda _{c}$ the force flips its sign (tending to implode the sphere).

It is worth mentioning that the Casimir effect for a sphere dates back to H. B. Casimir himself, who proposed that vacuum fluctuations might cause a conducting spherical shell to attract itself, in a way analogous to the case of two conducting plates. This conclusion was incorrect, as Boyer showed in 1968, since the Casimir stress in a perfectly conducting spherical shell is repulsive \cite{Boyer}. In this paper we showed that Lorentz violation, as described by the scalar field theory defined by the Lagrangian density (\ref{Lagrangian1}), allows us to tune the sign of the Casimir stress for a sphere in $D$-dimensions for $D>2$. In the three-dimensional case $D=3$, the critical value is $\lambda _{c} \approx 0.0025$. From a high-energy physics perspective, $\lambda _{c}$ is far from the allowed values for this parameter, since Lorentz symmetry violation is expected to be small. However, there are scenarios in which fundamental symmetries are naturally broken, e.g. condensed matter systems. Indeed, as shown in Ref. \cite{Grushin}, a condensed matter realization of a Lorentz-violating quantum electrodynamics (as described by the Standard-Model Extension) is the Weyl semimetal phase. This bridge between high-energy and condensed matter offers an opportunity to test phenomena initially predicted to occur at high-energies in solid state systems.

The present work can be extended in relevant ways. For example, the first task would be to incorporate the LV angular spacelike case, where in three-dimensions would correspond to add LV contributions to the $\hat{\theta}$- and $\hat{\phi}$-directions \footnote{Work in progress. These cases represent a different problem from the calculation and the physical point of view.}. Besides, since any realistic setup is necessarily immersed in a bath with a nonzero temperature, it would be interesting to determine the effects that thermal fluctuations would have in the Casimir stress. Studies in different geometries, as the cylindrical one, would also be of great value. We leave these problems for future works.


\acknowledgements

A. M.-R. acknowledges support from DGAPA-UNAM Project No. IA101320. C. A. E. is supported by a UNAM- DGAPA postdoctoral fellowship and Project PAPIIT No. IN111518. O. J. F. acknowledges support from DGAPA-UNAM Project No. IN103319. A.M.E-R. is supported in part by CONACyT grant 237351 (Mexico).

\appendix

\section{$c_k$ coefficients}
\label{AP1}

The lowest order coefficients $c _{k}$ appearing in Eq. (\ref{Exp1}) are

\begin{align}
c _{1} &= \frac{1}{\sqrt{\Lambda _{R}}} , \notag \\ c _{2} &= \frac{D ^{2} -3 D+2}{2 \sqrt{\Lambda _{R}}} , \notag  \\ c _{3} &= \frac{24 D ^{4} - 176 D ^{3} + 504 D ^{2} - 688 D + 3 \Lambda _{R} + 384}{192 \sqrt{\Lambda _{R}}} , \notag \\ c _{4} &= \frac{\left(D ^{2} - 5 D + 6 \right) \left(8 D ^{4} - 64 D ^{3} + 216 D ^{2} - 368 D + 3 \Lambda _{R} + 256 \right)}{384 \sqrt{\Lambda _{R}}} , \notag \\ c _{5} &= \frac{1}{737280 \sqrt{\Lambda _{R}}} \bigg[ 1920 D ^{8} - 38400 D ^{7} + 344320 D ^{6} - 1818624 D ^{5} + 160 D ^{4} (9 \Lambda _{R} + 38716) \notag \\ & \phantom{=} - 960 D ^{3} (17 \Lambda _{R} + 14456) + 160 D ^{2}(423 \Lambda _{R} + 123856) - 192 D (635 \Lambda _{R} + 85088) + 45 \left( - 35 \Lambda _{R} ^{2} + 1792 \Lambda _{R} + 131072 \right)\bigg] , \notag \\ c _{6} &= \frac{ \left(D ^{2} - 7 D + 10 \right)}{1474560 \sqrt{\Lambda _{R}}} \bigg[ 384 D ^{8} - 8192 D ^{7} + 79616 D ^{6} - 461312 D ^{5} + 160 D ^{4} (3 \Lambda _{R} + 10844 ) - 128 D ^{3} ( 45 \Lambda _{R} + 33556 ) \notag \\  & \phantom{=} + 32 D ^{2} ( 795 \Lambda _{R} + 210832 ) - 576 D ( 85 \Lambda _{R} + 10528 ) + 9 \left( - 175 \Lambda _{R} ^{2} + 3840 \Lambda _{R} + 262144 \right) \bigg] ,  \notag \\ c _{7} &= \frac{1}{1486356480 \sqrt{\Lambda _{R}}} \bigg[ 32256 D ^{12} - 1225728 D ^{11} + 21514752 D ^{10} - 231390208  D ^{9} + 20160 D ^{8} (3 \Lambda _{R} + 84344 ) \notag \\ &\phantom{=}  - 768 D ^{7} ( 2205 \Lambda _{R} + 11700884 ) + 8064 D ^{6} (2515 \Lambda _{R} + 4330708 ) - 21504 D ^{5} ( 6309 \Lambda _{R} + 4656325 ) \notag \\ & \phantom{=}  - 252 D ^{4} \left( 1575 \Lambda _{R} ^{2} - 2199920 \Lambda _{R} - 829993472 \right) + 280 D ^{3} \left( 21735 \Lambda _{R} ^{2} - 5060448 \Lambda _{R} - 1102971392 \right) \notag \\ & \phantom{=} - 252 D ^{2} \left( 127575 \Lambda _{R} ^{2} - 8778560 \Lambda _{R} - 1208414208 \right) + 72 D \left( 951825 \Lambda _{R} ^{2} - 26796672 \Lambda _{R} - 2490761216 \right) \notag \\ &\phantom{=} + 2835 \left( 565 \Lambda _{R} ^{3} - 17920 \Lambda _{R} ^{2} + 253952 \Lambda _{R} + 16777216 \right)\bigg] , \notag \\ c _{8} &= \frac{\left( D ^{2} - 9 D + 14 \right)}{2972712960 \sqrt{\Lambda _{R}}} \bigg[ 4608 D ^{12} - 184320 D ^{11} + 3426816  D ^{10} - 39236608 D ^{9} + 192 D ^{8} ( 63 \Lambda _{R} + 1604392 ) \notag \\ & \phantom{=} - 1536 D ^{7} ( 231 \Lambda _{R} + 1134236 ) + 2688 D ^{6} (1659 \Lambda _{R} + 2694956 ) - 2304 D ^{5} ( 13601 \Lambda _{R} + 9626180 ) \notag \\ &\phantom{=} - 36 D ^{4} \left( 3675 \Lambda _{R} ^{2} - 3737104 \Lambda _{R} - 1369374208 \right) + 64 D ^{3} \left( 33075 \Lambda _{R} ^{2} - 5640516 \Lambda _{R} - 1204408256 \right) \notag \\ & \phantom{=} - 36 D ^{2} \left( 319725 \Lambda _{R} ^{2} - 16406208 \Lambda _{R} - 2222780416 \right) + 72 D \left( 349125 \Lambda _{R} ^{2} - 7475328 \Lambda _{R} - 684670976 \right) \notag \\ & \phantom{=} + 405 \left( 3955 \Lambda _{R} ^{3} - 47040 \Lambda _{R} ^{2} + 516096 \Lambda _{R} + 33554432 \right)\bigg] .
\end{align}


\begin{thebibliography}{99}


\bibitem{Casimir} H. B. Casimir, Proc. K. Ned. Akad. Wet. \textbf{51}, 793 (1948).

\bibitem{Milonni}P. W. Milonni, \textit{The Quantum Vacuum: An Introduction to Quantum Electrodynamics} (Academic; New York, 2013).

\bibitem{Dalvit-Milonni-Roberts-Rosa}D. A. R. Dalvit, P. W. Milonni, D. Roberts and F. S. S. Rosa, \textit{Casimir Physics}, Lecture Notes in Physics (Springer-Verlag, Heidelberg, 2011).

\bibitem{Woods-et-al}L. M. Woods, D. A. R. Dalvit, A. Tkatchenko, P. Rodriguez-Lopez, A. W. Rodriguez and R. Podgornik, Rev. Mod. Phys. \textbf{88}, 045003 (2016).

\bibitem{Casimir-Polder}H. B. G. Casimir and D. Polder, Phys. Rev. \textbf{73}, 360 (1948).


\bibitem{Lamoreaux} S. Lamoreaux, Phys. Rev. Lett. \textbf{78}, 5 (1997); [Erratum: Phys.Rev.Lett. \textbf{81} 5475 (1998)].

\bibitem{Mohideen} U. Mohideen and A. Roy, Phys. Rev. Lett. \textbf{81}, 4549 (1998).

\bibitem{Roy} A. Roy, C.-Y. Lin and U. Mohideen, Phys. Rev. D \textbf{60}, 111101 (1999).

\bibitem{Klimchitskaya} G. Klimchitskaya and V. Mostepanenko, Mod. Phys. Lett. A \textbf{35} (03) (2020) 2040007.

\bibitem{Deutsch-Candelas} D. Deutsch and P. Candelas, Phys. Rev. D \textbf{20}, 3063 (1979).

\bibitem{DeRaad-Milton} L. L. DeRaad Jr. and K. A. Milton, Ann. Phys. (N.Y.) \textbf{136}, 229 (1981).

\bibitem{Dowker} J. Dowker and R. Banach, J. Phys. A \textbf{11}, 2255 (1978).

\bibitem{Aliev} A. N. Aliev, Phys. Rev. D \textbf{55}, 3903 (1997).

\bibitem{Huang} W.-H. Huang, Annals Phys. \textbf{254} 69  (1997).

\bibitem{Boyer} T. H. Boyer, Phys. Rev. \textbf{174}, 1764 (1968).

\bibitem{Milton 1} K. A. Milton, Ann. Phys. (N.Y.) \textbf{127}, 49 (1980).

\bibitem{Milton's book} K. A. Milton, \textit{The Casimir effect: Physical Manifestation of Zero-Point Energy} (World Scientific, Singapore, 2001).

\bibitem{Milton 2}  K. A. Milton, \textit{Casimir Energy for a Spherical Cavity in a Dielectric: Toward a Model for Sonoluminescence?} In: Bordag M. (eds) Quantum Field Theory Under the Influence of External Conditions. TEUBNER-TEXTE zur Physik, vol 30. Vieweg+Teubner Verlag, Wiesbaden (1996).

\bibitem{Johnson} K. Johnson. In B. Margolis and D. G. Stairs, editors, Particles and Fields 1979, p. 353, New York, 1980. AIR.

\bibitem{F-G-K} P. M. Fishbane, S. G. Gasiorowicz and P. Kaus, Phys. Rev. D \textbf{37}, 2623 (1988).

\bibitem{Graham 1} N. Graham, R. L. Jaffe, M. Quandt and H. Weigel, Phys. Rev. Lett. \textbf{87}, 131601 (2001).

\bibitem{Graham 2} N. Graham, R. Jaffe, V. Khemani, M. Quandt, M. Scandurra and H. Weigel, Nucl. Phys. B \textbf{645}, 49 (2002).

\bibitem{Graham 3} N. Graham, R. L. Jaffe and H. Weigel, Int. J. Mod. Phys. A \textbf{17}, 846 (2002).

\bibitem{Kostelecky 3}  D. Colladay and V. A. Kostelecky, Phys. Rev. D \textbf{55}, 6760 (1997).

\bibitem{SME2} D. Colladay and V. A. Kostelecky, Phys. Rev. D  \textbf{58}, 116002 (1998).

\bibitem{Finsler} B. R. Edwards and V. A. Kostelecky, Phys. Lett. B \textbf{786}, 319 (2018).

\bibitem{KFinsler} V. A. Kostelecky, Phys. Lett. B \textbf{701}, 137 (2011).

\bibitem{Cruz 1} M. B. Cruz, E. R. B. de Mello and A. Y. Petrov, Phys. Rev. D \textbf{96} 045019 (2017).

\bibitem{Cruz 2} M. Cruz, E. Bezerra De Mello and A. Y. Petrov, Mod. Phys. Lett. A \textbf{33}, 1850115 (2018).

\bibitem{Escobar-Medel-Martin} C. A. Escobar, Leonardo Medel and A. Mart\'in-Ruiz, Phys. Rev. D \textbf{101}, 095011 (2020).

\bibitem{PLB} C. A. Escobar, A. Mart\'{i}n-Ruiz, O. J. Franca and M. A. G. Garc\'{i}a, Phys. Lett. B {\bf 807}, 135567 (2020).

\bibitem{Gomes-Petrov} M. Gomes, J. R. Nascimento, A. Yu. Petrov and A. J. da Silva, Phys. Rev. D \textbf{81}, 045018 (2010).

\bibitem{Bender-Milton} C. M. Bender and K. A. Milton, Phys. Rev. D \textbf{50}, 6547 (1994).

\bibitem{Peskin} M. E. Peskin, D. V. Schroeder, \textit{An Introduction to quantum field theory}, Addison-Wesley, Reading, USA, 1995.

\bibitem{Abramowitz} M. Abramowitz and I. Stegun, \emph{Handbook of Mathematical Functions with Formulas, Graphs and Mathematical Tables}. Dover Publications. New York, 1972.

\bibitem{Milton 3} K. A. Milton, Phys. Rev. D \textbf{68}, 065020 (2003).

\bibitem{TIs} X.-L. Qi and S.-C. Zhang, Rev. Mod. Phys. \textbf{83}, 1057 (2011).

\bibitem{Wilczek} F. Wilczek, Phys. Rev. Lett. \textbf{58}, 1799 (1987).


\bibitem{Grushin} A. G. Grushin, Phys. Rev. D \textbf{86}, 045001 (2012).

\bibitem{CME} Q. Li, D. E. Kharzeev, C. Zhang, Y. Huang, I. Pletikosi\'{c}, A. V. Fedorov, R. D. Zhong, J. A. Schneeloch, G. D. Gu and T. Valla, Nature Physics \textbf{12}, 550 (2016).

\bibitem{Dzyaloshinskii} I. Dzyaloshinskii, E. M. Lifshitz and L. P. Pitaevskii, Adv. Phys. \textbf{10}, 165 (1961).

\bibitem{Cortijo} A. G. Grushin and A. Cortijo, Phys. Rev. Lett. \textbf{106}, 020403 (2011).

\bibitem{MCU} A. Mart\'{i}n-Ruiz, M. Cambiaso and L. F. Urrutia, Europhysics Lett. \textbf{113}, 60005 (2016).

\bibitem{Wilson} J. H. Wilson, A. A. Allocca and V. Galitski, Phys. Rev. B \textbf{91}, 235115 (2015).


\end{thebibliography}
\end{document}